\documentclass{aastex}
\usepackage[onecolumn]{emulateapj5}
\newcommand{\Tin}{T_{\rm in}}
\newcommand{\Rin}{R_{\rm in}}

\newcommand{\rg}{r_{\rm g}}

\slugcomment{2003/6/16 draft accepted version!}
\shorttitle{Watarai \& Mineshige}
\shortauthors{Model for GRS 1915+105: Variable Inner Edge }

\begin{document}
\title{Model for Relaxation Oscillations of Luminous Accretion Disk 
in GRS 1915+105: Variable Inner Edge}

\author{Ken-ya Watarai \altaffilmark{1,2,3}, 
        Shin Mineshige \altaffilmark{1}
}
\affil{1 Yukawa Institute for Theoretical Physics, Kyoto University,
        Sakyo-ku, Kyoto 606-8502}
\affil{2 Department of Astronomy, Graduate School of Science,
Kyoto University, Sakyo-ku, Kyoto 606-8502}
\affil{3 Research Fellow of the Japan Society for the Promotion of Science}

\email{watarai@yukawa.kyoto-u.ac.jp, watarai@kusastro.kyoto-u.ac.jp}

\begin{abstract}
To understand the bursting behavior of the microquasar GRS 1915+105,
we calculate time evolution of a luminous, optically thick
 accretion disk around a stellar mass black hole 
 undergoing limit-cycle oscillations
 between the high- and low- luminosity states. 
We, especially, carefully solve the behavior of the innermost part of
 the disk, 
since it produces significant number of photons during the burst,
and fit the theoretical spectra with the multi-color disk model.
The fitting parameters are $\Tin$ (the maximum disk temperature)
and $\Rin$ (the innermost radius of the disk).
We find an abrupt, transient increase in $\Tin$
and a temporary decrease in $\Rin$ during a burst,
which are actually observed in GRS 1915+105.
The precise behavior is subject to the viscosity prescription.
We prescribe the radial-azimuthal component of viscosity stress
 tensor to be $T_{r \varphi}=-\alpha \Pi (p_{\rm gas}/p)^{\mu}$,
with $\Pi$ being the height integrated pressure,
$\alpha$ and $\mu$ being the parameter, and $p$ and $p_{\rm gas}$ being
 the total pressure and gas pressure on the equatorial plane,
 respectively.
Model with $\mu=0.1$ can produce the overall time changes of
 $\Tin$ and $\Rin$,
but cannot give an excellent fit to the observed amplitudes.
Model with $\mu=0.2$, on the other hand, gives the right amplitudes,
but the changes of $\Tin$ and $\Rin$ are smaller.
Although precise matching is left as future work,
we may conclude that
the basic properties of the bursts of GRS 1915+105 can be explained by
our ``limit-cycle oscillation'' model.
It is then required that the spectral hardening factor
at high luminosities should be about 3 
at around the Eddington luminosity instead of less than 2 as is usually
 assumed. 
\end{abstract}


\section{Introduction}

Microquasars in our Galaxy are enigmatic objects on account of their
curious time variability, extremely large brightness (Belloni et
al. 1997a, 1997b; Mirabel et al. 1994; Rao, Yadav, \& Paul 2000),
 and energetic super-luminal jets (Mirabel \& Rodriguez 1998). 
Large numbers of broad-band observations have been so far performed,
revealing that
they are low mass X-ray binaries including black holes and that 
the accretion flow around the black hole illuminates via the release of
the gravitational energy. 
Their long-term bursts ($\sim$ months) could be triggered by a thermal
instability in the disk (e.g. Mineshige \& Wheeler 1989), however, 
the mechanism causing their short-term peculiar behavior is not well
understood yet (Mirabel \& Rodriguez 1999; Belloni et al. 2000; Naik et al. 2002).
From the theoretical viewpoints, it is interesting to investigate how
their time-dependent properties of accretion disks can be understood in
the frame work of the disk models. 

Among plenty of Galactic black-hole candidates, GRS 1915+105, most
famous microquasar, 
is unique in exhibiting quasi-periodic luminosity variation
 (Belloni et al. 1997a, 1997b). 
Yamaoka, Ueda \& Inoue (2001) analyzed the X-ray data of GRS 1915+105,
 taken by ASCA and RXTE during 1998-2000 and found that it 
exhibits mainly two branches in its high luminosity state: 
the high temperature branch (HTB), $kT\sim 2$ keV, and the low
temperature branch (LTB), $kT \lesssim 1$ keV
 (see also Belloni 2002). 
The transition time between them is $\sim 1$ s. Interestingly,
both states can be fitted by the multi-color disk model
 (hereafter, MCD; Mitsuda et al. 1984),
 however, what is puzzling in their HTB data is
 too high temperature and too small inner radius ($\Rin \sim 20$ km)
 to account for within the framework of the standard disk model
 for black-hole candidates (BHCs). 
It is interesting to note, in this respect,
 that a similar puzzle is known
 in ultra-luminous compact X-ray sources
 (ULXs, see Makishima et al. 2000, Mizuno, Kubota \& Makishima 2001).
In this case, several possibilities have been discussed;
rotating black-hole hypothesis (Makishima et al. 2000),
moderate beaming effects (King et al. 2001), 
and the shift of the inner edge of the disk in super-critical accretion
regimes (Watarai, Mizuno, \& Mineshige 2001, hereafter WMM01).  

Then, why is GRS1915+105 so unique? 
What parameter is peculiar to GRS 1915+105?
The fact that its luminosity is very large,
 comparable to the Eddington luminosity, indicates that
 this source seems to suffer very high mass accretion flow.
The accretion disk theory predicts that
 an accretion disk becomes secularly and thermally unstable,
 when the mass accretion rate is very high so that the disk
 is radiation pressure-dominated
 (Lightman \& Eardley 1974; Shibazaki \& H$\rm{\bar{o}}$shi 1975; 
Shakura \& Sunyaev 1976).
What are then the observable consequences of
 such an unstable accretion disk?

As for the solution of accretion flow at accretion rates
exceeding the critical rate ($\dot{M}_{\rm crit}=L_{\rm E}/c^2$),
Abramowicz et al. (1988) found a self-consistent, steady-state solution,
which is now known as the ``slim disk.''  
It was then suggested that a near-critical disk may undergo
limit-cycle oscillations, like the case of dwarf-nova outbursts,
between the gas pressure-dominated standard disk and the slim disk
(Kato 1983; Abramowicz, Lasota, \& Xu 1986).
Honma, Matsumoto \& Kato (1991, hereafter HMK91) were the first to
 numerically demonstrate limit-cycle oscillations between the two
 different states based on the slim disk model. 
 They show that an existence of the enough radiation
 pressure dominated region is essential for developing the thermal
 instability. 
Szuszkiewicz \& Miller (1997, 1998) confirmed
 their results for wider $\alpha$ parameter ranges. 
The X-ray spectral variability of these oscillations was calculated by
Zampieri, Turolla, \& Szuszkiewicz (2001), who discussed the spectral
 behavior in each state and the corresponding hardening factor. 
More recently, Szuszkiewicz \& Miller (2001) calculated the non-linear
evolution of the disk with diffusion type viscosity, 
which is a more physical, and
could also reproduce limit-cycle oscillation as in previous papers. 

In the present study, we pay special attention to the behavior of the
innermost region. 
The steady calculation of the slim disk indicates that
 a super-critical accretion flow can emit substantially from the
part even inside $3\rg$, since large amount of gas exists inside $3\rg$
because of high mass accretion flow.
 Therefore, it has been anticipated to see a temporary reduction of
the radius of the inner edge of the disk during a burst
 (Watarai et al. 2000; Mineshige et al. 2000).
 To examine if this is the case,
 we carefully calculate the time evolution of the innermost part of a
 super-critical flow,
 compare the theoretical results with observations, and
discuss the possibility of obtaining a simpler proof for the existence
of the super-critical accretion and thus probing the disk-instability
to be a cause of the light variations of GRS 1915+105. 
Note that previous calculations were all performed 
down only to the radius of $\sim 2.7 \rg$ (as in HMK91) or $\sim 2.5
\rg$ (Szuszkiewicz \& Miller 1997, 1998, 2001; Zampieri et al. 2001) so
that an important fraction of the total radiation release might have
been missed.
%Szuszkiewicz and Miller (1997, 1998, 2001) calculated down to 2.5 $\rg$,
%however, the spectral properties of time-dependent disk were not shown
%in their paper.  
%Zampieri et al. (2001) compared the spectral behavior of the time
% dependent disk with the observation of BHCs,
% then they discussed that the inner boundary effect ($\sim 2.5 \rg$) is 
% able to increase .   
%
In the present study, 
 we calculate the disk structure down to $\sim$ 2.2 -- 2.4$\rg$,
 well inside the sonic radius, to see an observable feature
 in terms of the fitting parameters more clearly. 
This smaller boundary is critical 
to test the disk-instability scenario for GRS 1915+105.

In the next section we present the basic equations and
 our model assumptions. 
We show the results of the time-dependent calculations of the flow in
section 3. 
The comparison with the X-ray observation of the microquasar
 GRS 1915+105 is discussed in section 4. 
The last section is devoted to conclusions. 

\section{Basic Equations and Numerical Methods}

The methods of calculation are nearly the same as those by HMK91 (see
Kato, Fukue, and Mineshige 1998 for the detailed discussion on the basic
equations).  
The basic equations are written by using the cylindrical coordinates
 ($r,\varphi, z$). 
The assumptions of our calculation are; 
(i) a disk is axisymmetric, 
(ii) pseudo-Newtonian potential
 (Paczy\'{n}sky \& Wiita 1980) is adopted; that is, $\psi=-GM/(R-\rg)$
 with $R \equiv \sqrt{r^2+z^2}$, and 
(iii) we employ one zone approximation toward the vertical direction. 
Here, $\rg$ is the Schwarzschild radius defined by $2GM/c^2 = 3 \times
10^6 (M/10M_{\odot})$ cm. 
We use the height integrated quantities, such as 
\begin{equation}
\Sigma = \int_{-H}^{H} \rho dz = 2 I_{\rm N} \rho H,
\end{equation}
and 
\begin{equation}
\Pi = \int_{-H}^{H} p dz = 2 I_{\rm N+1} p H.
\end{equation}
Here, $\Sigma$, $\Pi$, $\rho$, $p$, and $H$ are surface density, height
integrated pressure, mass density, total pressure, and scale height,
 respectively. 
The coefficients, $I_{N}$ and $I_{N+1}$, are introduced by H$\bar{\rm o}$shi
(1977). 
The density and pressure are related to each other by the polytropic
relation, $p \propto \rho^{1+1/N}$. 
We assign $N=3$ in the entire calculations
 (i.e., $I_{3}=16/35$ and $I_{4}=128/315$). 
The equation of state is 
\begin{equation}
 p = p_{\rm rad} + p_{\rm gas} = \frac{a}{3}T_{c}^4 
+ \frac{k_{\rm B}}{\bar{\mu}m_{\rm H}}\rho T_c, 
\end{equation}
where the first term on the right hand side represents the radiation pressure
($a$ and $T_c$ are the radiation constant and the temperature 
on the equatorial plane, respectively) 
and the second term represents the gas pressure 
($k_{\rm B}$ is the Stefan-Boltzmann constant, 
 $\bar {\mu} = 0.5$ is the average molecular weight,
and $m_{\rm H}$ is the hydrogen mass). 
We assume hydrostatic equilibrium (H${\rm \bar{o}}$shi 1977), 
\begin{equation}
 (2N+3) \frac{\Pi}{\Sigma} =H^2 \Omega_{\rm K}^2
           \equiv H^2\frac{GM}{r(r-\rg)^2}.
\end{equation}
Here, $\Omega_{\rm K}$ denotes the Keplerian angular frequency
 under the pseudo-Newtonian potential. 
%$\Omega_{\rm K} = r/(r-\rg) (GM/r^3)^{1/2}$. 

Mass conservation, the radial component of the momentum conservation,
and angular momentum conservation are written as follows:
\begin{equation}
\frac{\partial}{\partial t}(r \Sigma) + 
\frac{\partial}{\partial r}(r \Sigma v_r) 
%+ \frac{\partial}{\partial z}(\rho v_z) 
= 0, 
\end{equation}
\begin{equation}
\frac{\partial}{\partial t}(r \Sigma v_r)  + 
\frac{\partial}{\partial r}(r \Sigma v_r^2) +
r \frac{\partial \Pi}{\partial r} + 
\Pi \frac{\partial {\rm ln} \Omega_{\rm K}}{\partial {\rm ln} r}
%+ \frac{\partial}{\partial z}(\rho v_r v_z) 
% r\frac{\Sigma v_{\varphi}^2}{r} 
= 
 \Sigma (l^2-l_{\rm K}^2), 
\end{equation}
and
\begin{equation}
\frac{\partial}{\partial t}(r^2 \Sigma v_{\varphi}) + 
\frac{\partial}{\partial r}(r^2 \Sigma v_r v_{\varphi}) 
%+ \frac{\partial}{\partial z}(\rho v_{\varphi} v_z)
%+ \frac{\rho v_r v_{\varphi}}{r} 
= 
\frac{\partial}{\partial r}(r^2 T_{r \varphi}),
\end{equation}
Here, the velocity is expressed by $v_r$ and $v_{\varphi}$ for the
radial and azimuthal components, respectively, and 
the angular momentum of gas is expressed by $l$ and $l_{\rm K}$, which
are defined by $l \equiv r v_{\varphi}$ 
and $l_{\rm K} \equiv r^2 \Omega_{\rm K}$, respectively.  

The $r$-$\varphi$ component of viscous stress tensor in
equation (7) is prescribed as 
\begin{equation}
T_{r \varphi} \equiv \int_{-H}^{H} t_{r \varphi} dz 
= -\alpha \beta^{\mu} \Pi, 
\label{trpvis}
\end{equation}
where $\mu$ is a parameter ($0 \le \mu \le 1$) and $\beta \equiv p_{\rm
gas}/p$ (Szuszkiewicz 1990; HMK91; Watarai \& Mineshige 2001). 
In the limit of vanishing $\mu$, we recover the ordinary $\alpha$ viscosity
prescription (Shakura \& Sunyaev 1973). 
If we take the limit of $\mu \to 1$, on the other hand, the viscosity
depends on the gas pressure only (Sakimoto \& Coroniti 1981). 

Finally, the energy equation is
\begin{equation}
\frac{\partial}{\partial t}(r \Sigma \epsilon_{\rm tot}) 
+ \frac{\partial}{\partial r}
[r(\Sigma \epsilon_{\rm tot}+\Pi)v_r -r T_{r\varphi}v_{\varphi}]=
-2 rF^-,
\end{equation}
in which advective cooling, viscous,
 and radiative cooling are considered. 
The explicit form of the total energy ($\epsilon_{\rm tot}$) is 
\begin{equation}
\epsilon_{\rm tot}=\left[3(1-\beta)+\frac{\beta}{\gamma-1} 
                  + \frac{1}{2}\right]\frac{\Pi}{\Sigma} 
                  + \frac{1}{2}(v_r^2+v_{\varphi}^2)+\psi_0(r),
\end{equation}
where the first term on the right-hand side is the internal energy of
gas ($\gamma$ is the adiabatic index and we set $\gamma=5/3$ 
in the present calculation),
the second term represents the kinetic energy, 
and the last term is the potential
energy on the equatorial plane, $\psi_0 (r) \equiv -GM/(r-\rg)$. 
Radiative cooling flux per unit surface area of optically thick medium
is given by  
\begin{equation}
  F^- = \frac{8acT_c^4}{3\tau},
\end{equation}
where, $\tau$ is the optical depth: 
\begin{equation}
\tau = \bar{\kappa} \Sigma = (\kappa_{\rm es} +  \kappa_{\rm ff})\Sigma, 
\end{equation}
where $\bar{\kappa}$ is average opacity, $\kappa_{\rm es}=0.4$ is
a opacity of the electron scattering, $\kappa_{\rm ff} = 0.64\times
10^{23} \bar{\rho}~\bar{T}^{-7/2}$ is the 
absorption opacity via thermal Bremsstrahlung, and $\bar{\rho}=16/35
\rho$ and $\bar{T}=2/3 T_c$ are
vertically averaged density and temperature, respectively. 
We solve equations (5)--(7) and (9) by modified Lax-Wendroff method with
artificial viscosity. 

To examine the behavior of the innermost part of the accretion flow, 
the calculations are performed from the outer radius at $2000 \rg$ down
to the inner radius
$\sim 2.2 \rg$ through the transonic point (Matsumoto et al. 1984). 
The number of total mesh is $400\sim 500$ and each mesh point is
distributed uniformly on the logarithmic scale at large radii, $r > 3.2 \rg$,
and also uniformly but on the linear scale at smaller radii,
where each mesh size is $\sim 0.01 \rg$. 
We fixed the black-hole mass to be $m=10$ ($m=M/M_\odot$, where
$M_\odot$ is solar mass) and 
the viscosity parameter to be $\alpha=0.1$. 
The critical accretion rate, $\dot{M}_{\rm crit}$, is defined by
$L_{\rm E}/c^2$, where $L_{\rm E}$ is the Eddington luminosity 
and $c$ is the speed of light. 
We define the dimensionless accretion rate to be
$\dot{m}\equiv\dot{M}/\dot{M}_{\rm crit} = \dot{M}c^2/L_{\rm E}$
 throughout the present study.

\section{Time Evolution of the Inner Disk}

\subsection{Bolometric Light Curves}
\begin{figure}[h]
%\figurenum{1}
\epsscale{0.8}
%\plotone{lc3.eps}
\plotone{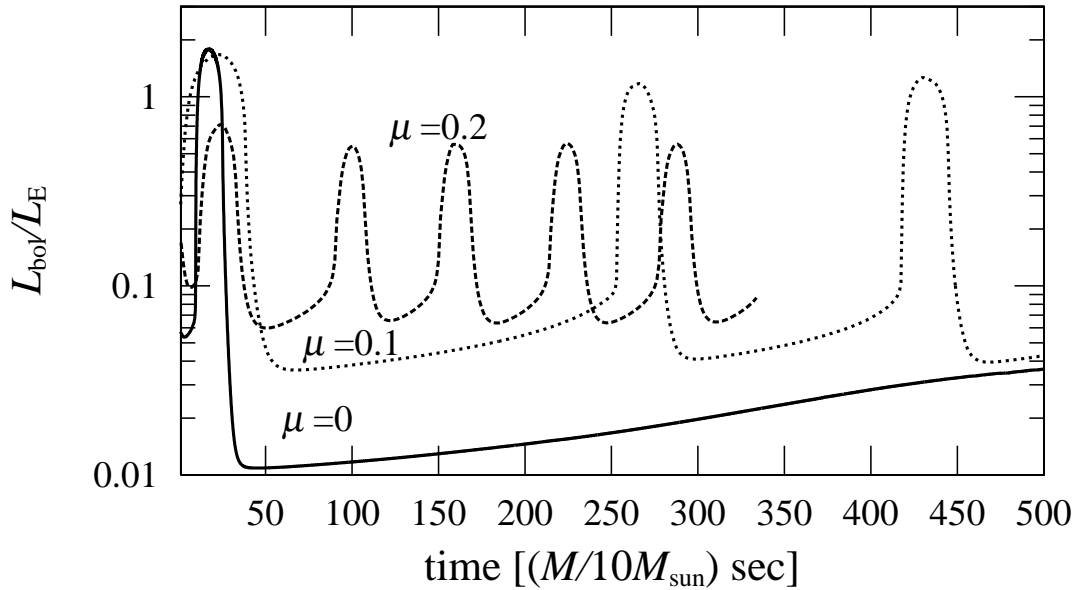}
\caption{Bolometric light curves for different $\mu$ parameters
 ($\mu$=0, 0.1 and 0.2).  The bolometric luminosity is defined by 
$L_{\rm bol} \equiv \int 2 \pi r^2 F^- dr$. 
The bolometric luminosity is normalized by the Eddington luminosity.
%The numbers in the figure 
%represent the time sequence numbers for the cases with
%$\mu = 0.0$ and $\mu = 0.2$, respectively (see Figs 2 and 3). 
%The open-square and the number with $'$ represents the case of $\mu=0.2$. 
}
\label{fig:lc}
\end{figure}

We first show the calculated light curves for models with
$\mu=0$ (by the solid curve), 0.1 (by dotted curve) and 0.2
 (by the dashed curve) in figure \ref{fig:lc}. 
The bolometric luminosity is calculated by integrating 
$2 \pi F^{-}(r)$ from the inner edge at $r=2.2\rg$
 (at $r=2.4\rg$ for $\mu=0.1$ case)
to the outer edge at $r = 2000 \rg$.
Eddington luminosity for a $10M_{\odot}$ black hole is 
$1.25 \times 10^{39}$ ($M/10M_{\odot}$) erg ${\rm s^{-1}}$. 
For simplicity,
we assume blackbody radiation field (see section 3.4 for corrections).

In model with $\mu=0$, the disk luminosity suddenly rises on
timescale of $\lesssim 5(M/10M_{\rm \odot})$s, 
keeps a high value for $\sim 10 (M/10M_{\rm \odot})$s, 
and then decays. 
These time scales roughly correspond to the viscous timescale, 
 $t_{\rm vis} \sim \alpha^{-1} \Omega^{-1} (r/H)^2$,
 in the inner region.  
Namely, 
\begin{equation}
 t_{\rm vis} \sim 1.6~\left(\frac{M}{10M_\odot} \right) \left(\frac{r}{5\rg} \right)^{3/2} \left(\frac{\alpha}{0.1} \right)^{-1}
 \left(\frac{H/r}{0.1} \right)^{-2} sec.
\end{equation}

The amplitude of the burst is about one order of magnitude. 
Here, we calculated the same model as that in HMK91, i.e., $\mu=0$,
$\dot{m}=0.96$, $m=10$, and $\alpha=0.1$. 
We confirmed that the amplitude and the duration %, $t_{\rm burst}$, 
of the burst in our calculation agree with those of HMK91 ($t_{\rm
burst} \sim 15$ second for this parameter set).  
We continued calculations until $\sim$ 2000 second (that is, 2$\times
10^7$ ($\rg/c$) calculation steps), however, we could
not obtain a next burst in the same parameters as those in HMK91.  
In our calculation, small oscillations (or small bursts) took place
 near to the critical accretion rate for many times.
Moreover, $\mu=0$ model takes a lot of recurrence time so that 
observational feature of GRS1915+105 does not agree with $\mu=0$. 
Hence, we will show $\mu=0.1$ model instead of $\mu=0$ model
 in the next section. 

In model with $\mu=0.2$, in contrast, 
the amplitude of the burst is less, only about $\sim$ 2 mag
(about a factor of $\sim 6$ variations). 
The burst rise time is somewhat longer than in the previous case 
but not significantly.  
%and the burst duration is $\sim$ 20 s. 
Both of the bolometric light curves look quite similar to 
those obtained by HMK91,
who also found that the burst amplitude and time scale 
are sensitive to the parameter, $\mu$. 
The different burst amplitudes can be understood in terms of
the different shapes of the S-shaped thermal equilibrium curves,
since the ranges of surface density (and thus of effective temperature),
in which solutions are unstable, shrinks with increasing $\mu$ (HMK91).
We will explain detailed evolution in the next two subsections. 

%%%%%%%%%%%%%%%%%
%Fig. 1
%\begin{figure}[p]
%%%%%%%%%%%%%%%%%%%

\subsection{State transition: case of $\mu=0.1$}

As we mentioned in previous subsection, we can not obtain perfectly
relaxed solutions in case of $\mu=0$ so that we will show $\mu=0.1$
case.  
We apply the data at third burst in figure \ref{fig:lc}, 
because the first burst is affected by initial conditions.

\begin{figure}[h]
%\figurenum{1}
\epsscale{0.7}
\plotone{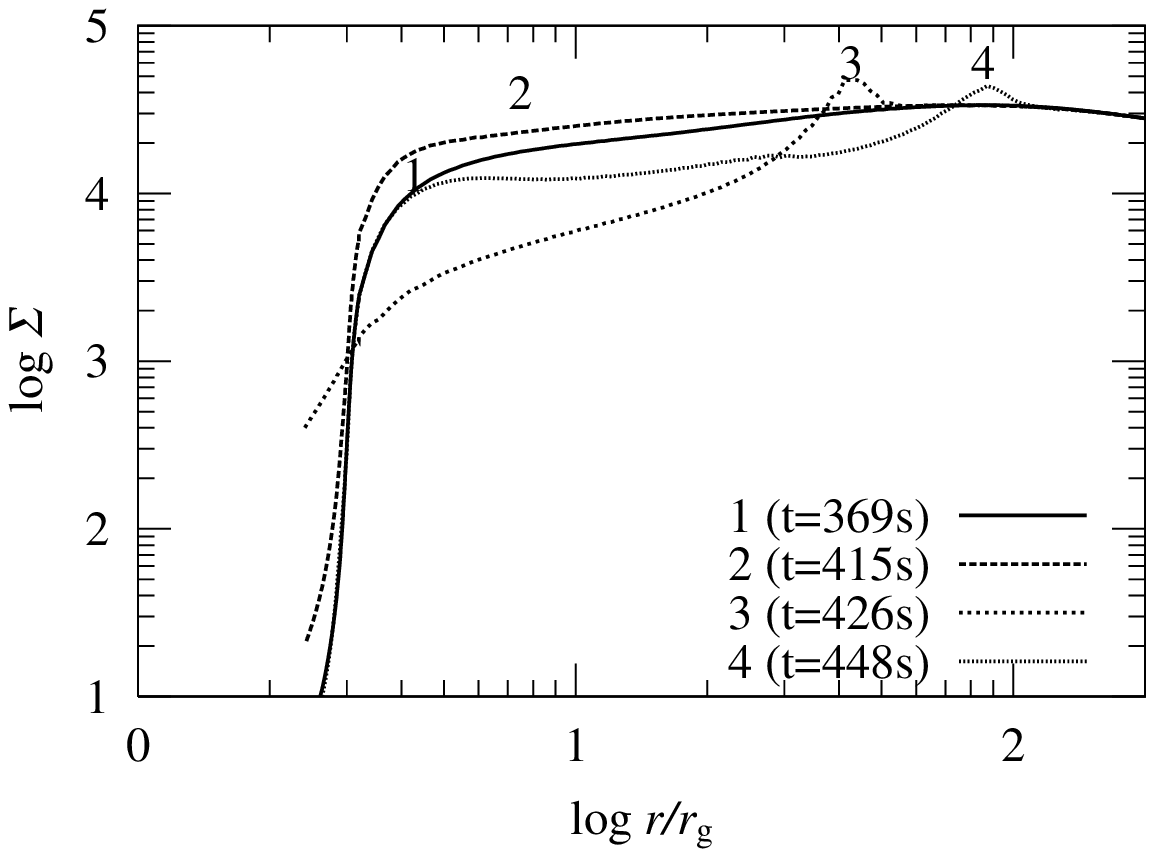}
\plotone{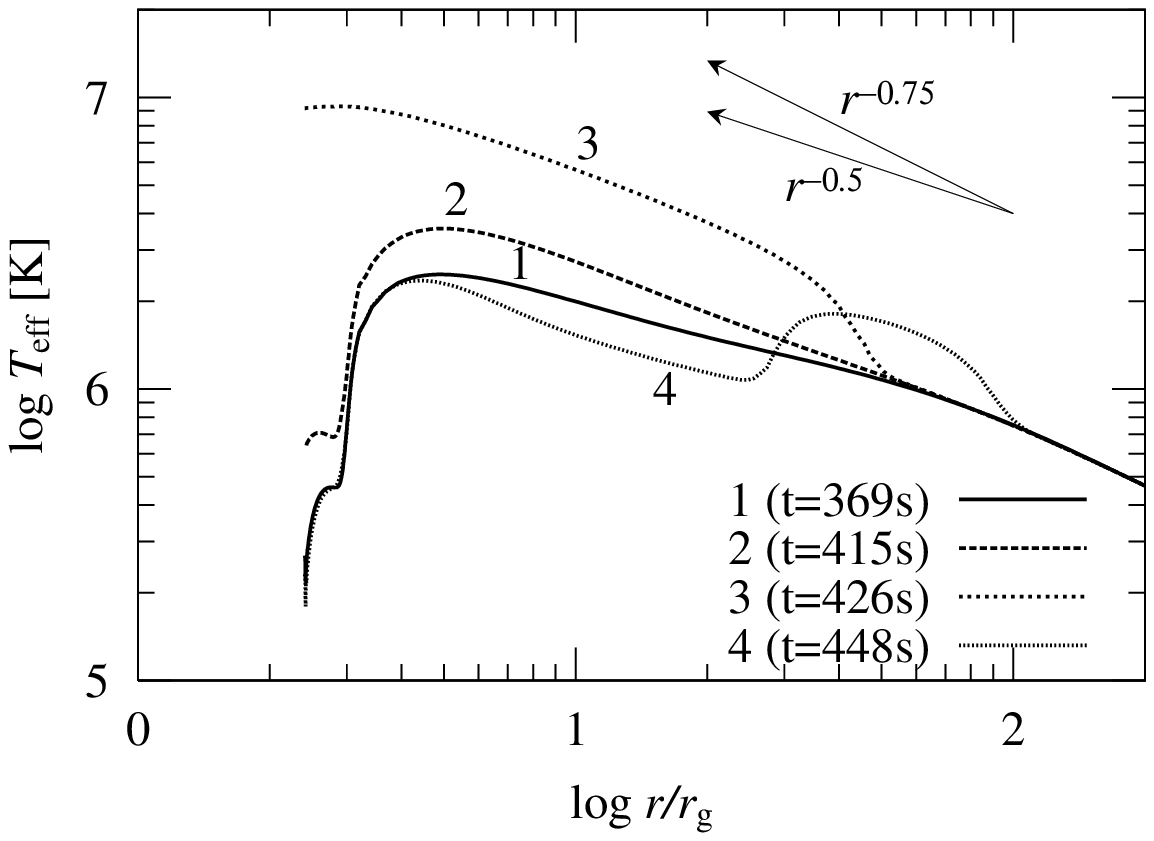}
\caption{Time evolution of the surface-density profile (top)
 and the effective temperature profile (bottom)
 for the case with $\mu=0.1$. 
The numbers 1 -- 4 indicate the time sequence numbers 
(see figure \ref{fig:lc}).
 The solid curve labeled with the number 1 
represents prior burst profile.
The thick dashed lines (time sequence 2), dotted lines 
(time sequence 3), and small dashed lines represent the profiles in
 the rise, peak, and decay phases, respectively. 
% and each time sequence is separated by the constant interval of 
%$4\times 10^5 ~(\rg/c)=40 (M/10 M_{\odot})$ s. 
Black-hole mass and the viscous parameter are 
$M=10M_{\odot}$ and $\alpha=0.1$. 
We assume a constant mass input at a rate of $\dot{m}_{\rm in}=3.0$.
} 
\label{fig:tevol1}
\end{figure}

Figure \ref{fig:tevol1} illustrates
the evolution of the radial distributions of
the surface density (top) and effective temperature (bottom).
%We set mass input rate of $\dot{m}_{\rm in}=0.96$ and $\mu=0$ 
%(i.e., the ordinary $\alpha$ viscosity prescription is adopted). 
We set mass input rate of $\dot{m}_{\rm in}=3.0$ and $\mu=0.1$. 
 The numbers in each panel indicate the order of the evolution sequence
from third burst of $\mu=0.1$ model in figure \ref{fig:lc}.

The solid curve indicated by the number `1'
represents the state before the burst. 
In the prior burst state, radiation pressure exceeds the gas pressure from $\sim 5 \rg$ to $20 \rg$.

Initially, there is little mass inside $3 \rg$
(this radius corresponds to the radius of the marginally stable 
circular orbit around a non-rotating black hole). 
Surface density abruptly drops by 4 orders of magnitude inside $3 \rg$ 
in the initial state, because the radial velocity of the gas prominently
increases near the transonic point. 
After $\sim$ 5$(M/10M_{\rm \odot})$s, a burst starts.  
A thermal instability is ignited in the innermost region
and an upward transition from the radiation pressure-dominated
standard-disk branch to the slim-disk one is initiated there. 
Then, a transition wave propagates outward and the outer zone
subsequently undergoes an upward transition.
Once the instability sets out, 
the efficiency of angular momentum extraction becomes largely enhanced, 
thus, gas in the disk being able to fall into the black-hole rapidly. 
Thus, surface density in the inner part shows a rapid decline.
This enhanced mass flow fills the inner empty zone, and, hence,
the surface density inside $3 \rg$ suddenly increases 
by more than one order of magnitude.

%At the peak (corresponding to the time sequence number 3),
%the surface density profile is roughly $\Sigma \propto r^{1/2}$
%and the radial velocity has a dependence of $v_r \propto r^{-3/2}$. 
%These profiles are in good agreement with those of 
%the self-similar solutions for
%the steady, slim disk (see Watarai 2002, in preparation). 

The effective temperature distribution also shows interesting features. 
The most important feature seen around the peak is that 
because of enhanced mass flow toward the innermost part
the temperature inside $3 \rg$ keeps a rather high value, $> 10^7$ K,
so that significant amount of radiation is now expected from there. 
This property can be seen in figure 11 of Abramowicz et al. (1988)
and in figure 2 of Watarai et al. (2000) have explicitly demonstrated
that this is observable. 
In the high, slim-disk state, the maximum temperature is
slightly higher and the inner disk radius becomes smaller
than in the low, standard-disk state.
The radial distribution of the effective temperature in time sequence
numbers 3 are proportional to $r^{-1/2}$, 
which is in good agreement with the analytical
prediction (Watarai \& Fukue 1999; Wang \& Zhou 1999). 
We plotted $r^{-1/2},~r^{-3/4}$ slopes in figure 2b, 3b for reference. 

Another notable feature is found in the subsequent evolution. 
When the instability wave reaches $\sim 100 \rg$, 
a downward transition from the slim-disk branch to the standard-disk one
begins from the inner region and again propagates outward.
In both phases the transition is initiated from the inner region and the
instability is propagated outward. 
Although these features are basically the same as those of the 
previous works (HMK91), we find interesting features
in the behavior of the innermost region, which has been unclear in
the previous calculations because of the larger truncation radius.
This produces interesting observable effects. 
 It should be noted that effective optical thickness at the inner
region is more than unity during the burst phase. 
Hence, we can expect thermal emission around the disk inner region.

On the other hand, we have confirmed that a relatively large viscosity
parameter, $\alpha \sim 0.1$, is required to reproduce the observations
of GRS1915+105.   
If we adopt a smaller viscosity parameter, 
 which tends to increase surface density and thus optical depth,
 the duration becomes longer than the observed value.  
Therefore, we conclude that $\alpha$ is likely to be about 0.1 and some
additional sources, such as metals, may contribute and raise opacities.

%Fig. 2
%\begin{figure}[p]

\subsection{Different viscosity prescription: case of $\mu=0.2$}

\begin{figure}[h]
%\figurenum{2}
\epsscale{0.7}
\plotone{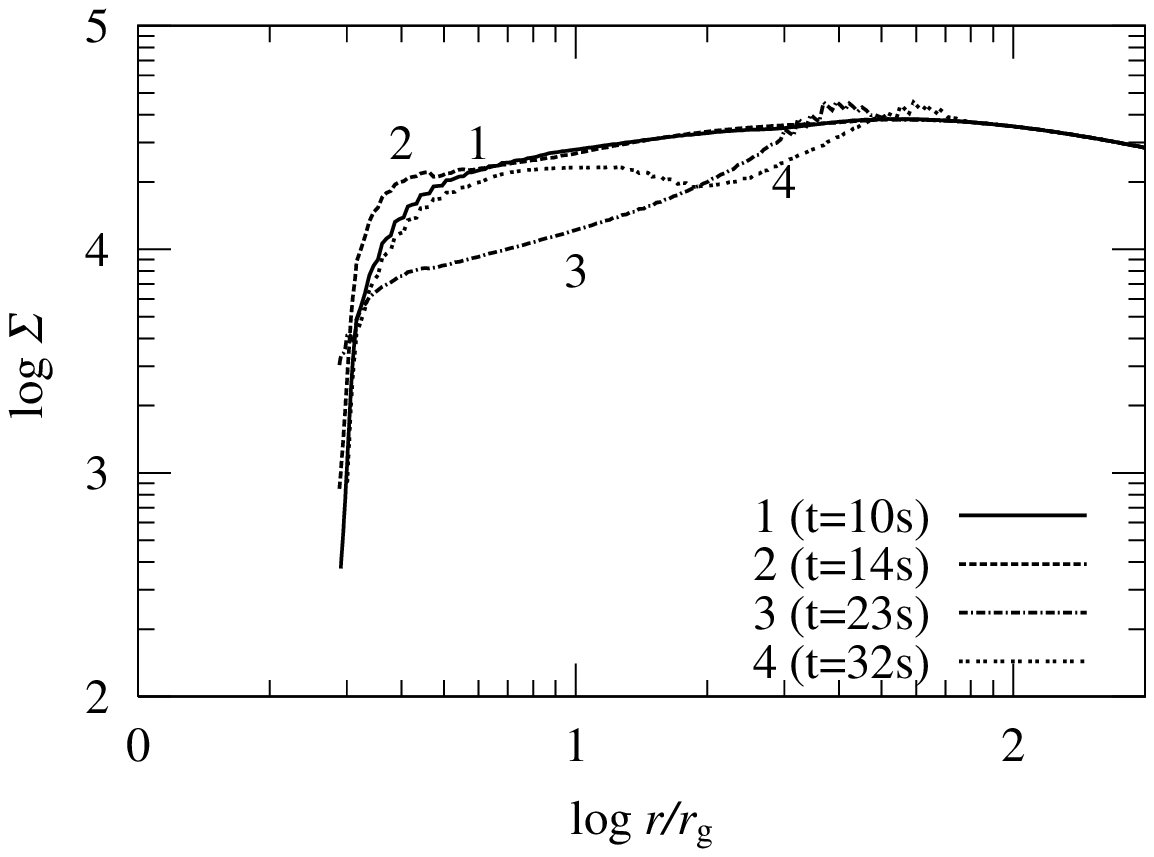}
\plotone{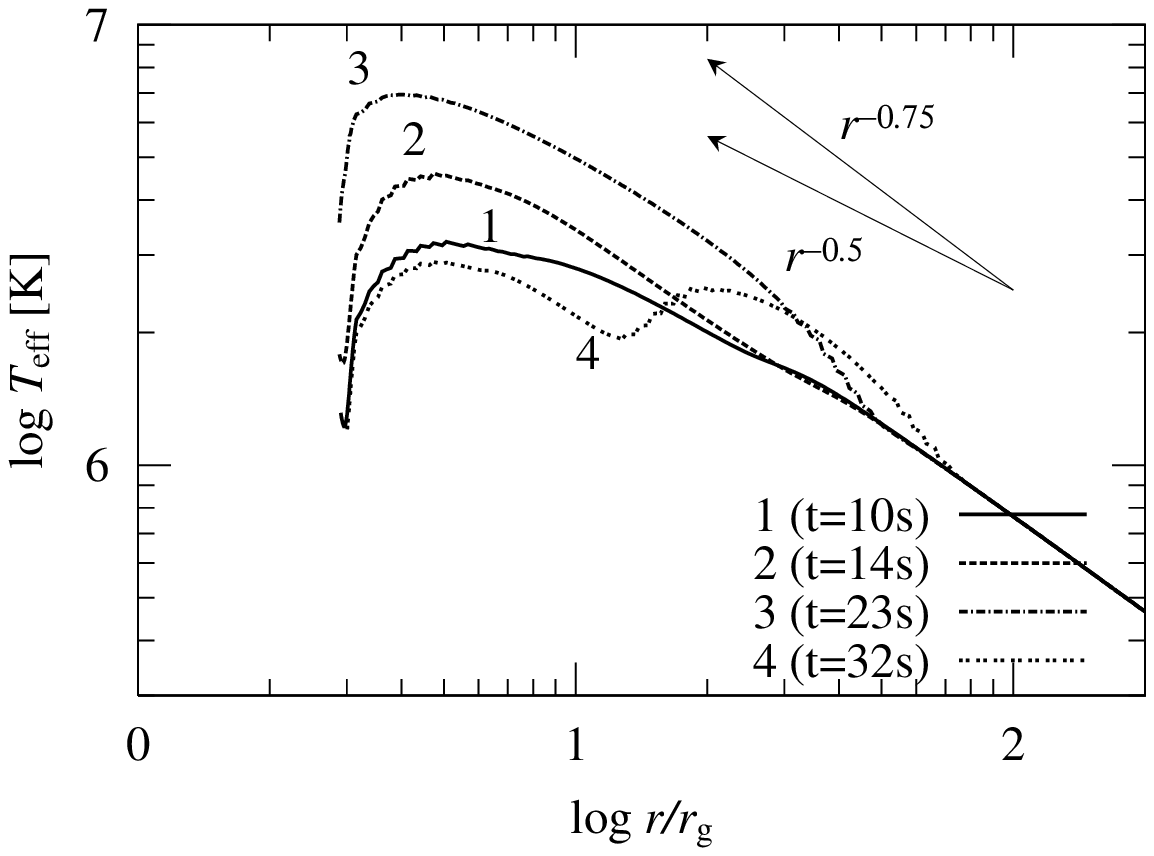}
\caption{The same as figure \ref{fig:tevol1} but for the case with
 $\mu=0.2$ (We prescribe the $r$-$\varphi$ component of the shear stress tensor
 to be $T_{r \varphi}=-\alpha \Pi (p_{\rm gas}/p)^{\mu}$).  
 We assume a constant mass input at a rate of $\dot{m}_{\rm in}=3.0$.
Other parameters are the same as those in figure \ref{fig:tevol1}. 
Note that even at the peak surface density abruptly decreases inside $3\rg$
and the temperature maximum is reached outside $3\rg$,
which makes a big contrast to the case with $\mu =0.1$.
}
\label{fig:mu02}
\end{figure}

Honma et al. (1991) used the viscosity prescriptions
given by equation (\ref{trpvis}) and discussed how different prescriptions
affect the overall disk evolution.
They have demonstrated that in the case with $\mu=0.25$
the burst duration is shorter and the amplitude is smaller 
both by a factor of $\sim 2$, compared with the case with $\mu=0$. 
That is, modifying viscosity prescription leads to drastic changes
in the amplitudes of light variations and in the burst duration. 
We also test the case with $\mu=0.2$ 
and display the resultant light variations in figure \ref{fig:lc}
and the structural evolution in figure \ref{fig:mu02}.
Radiation pressure dominates over gas pressure from $\sim 4 \rg$ to $65
\rg$ in the initial state ($\dot{m}_{\rm in}=3.0$).

Interestingly, some differences are found.
Although the surface density and the temperature of the innermost region
are still variable,
the way of time changes are qualitatively different from the previous case.
The upper panel of figure \ref{fig:mu02} indicates that the instability
evolves less efficiently than in the case with $\mu=0$.
Especially, surface density remains very low inside $3 \rg$ even at the peak
and the maximum temperature does not exceed $10^7$ K.  Moreover,
the maximum temperature is reached at $\sim 4 -6 \rg$
unlike the previous case, in which temperature is maximum at the inner edge.

The reason for this change can be understood as follows:
Non-zero $\mu$ values lead to a reduction of $\alpha$ when the radiation
pressure is larger than the gas pressure ($p_{\rm rad}>p_{\rm gas}$). 
%Then, both temperature and radiation flux should be less, if $\mu > 0$.
This reduction tends to increase the surface density of the
equilibrium solution at a fixed mass accretion rate
(since $\alpha\Sigma \propto {\dot M}$ in the steady condition),
and the amount of this increase gets larger, as $\dot M$ increases.
In other words, the S shape of the thermal equilibrium
curves get less prominent, as $\mu$ increases
(see Kato et al. 1998, figure 10.4). 
As a result, the thermal instability tends to be suppressed.

\subsection{Model Spectral Fitting}

We next perform the spectral fitting to the calculated spectra in the 
same manner as that adopted by WMM01.  
First, we calculate the spectra based on our calculation, assuming
blackbody radiation at each radius for each time step. 
Then, we fit the spectra with the MCD model (Mitsuda et al. 1984),
 from which we can derive the disk maximum temperature,
 $T_{\rm in}^0$, and the radius,
 $R_{\rm in}^0$, of the region emitting blackbody radiation 
 with $T_{\rm in}^0$.  
Before making a direct comparison with the observations
we need the following corrections:
(i) The derived value of $R_{\rm in}^0$ is influenced by
the effects of the inner boundary condition 
as well as general relativistic effects; 
that is, the value tends to be
overestimated, in general, compared with the real inner-edge radius. 
We thus need to introduce a correction factor $\eta=0.41$
and set $R_{\rm in}= \eta R_{\rm in}^0$ (Kubota et al. 1998).   
(ii) Second, the disk spectra are not perfectly blackbody but are
modified by the Compton scattering effects, which lead to
spectral hardening
 (Ross, Fabian, \& Mineshige 1992; Shimura \& Takahara 1995). 
Zampieri et al. (2001) also estimated the spectral hardening factor
through solving the radiation transfer in the vertical direction,
and found results, which mostly agree with Shimura \& Takahara (1995).
The actual value is $\kappa \approx 1.7$ for stellar mass black hole
and increases with increasing $\dot M$.

 Unfortunately, it is difficult to precisely predict the spectral
hardening factor due mainly to the uncertainties in metal opacities
(which depend on the metal ionization stages, amount of incident X-ray
photons, etc). 
Poorly understood disk-corona structure should also modify the model
spectra. 
Also it strongly depends on the line of sight of observation, and the
radial motion of the flow. 
Previous works (Shimura \& Takahara 1995; Wang et al. 1999;
 Zampieri et al.2001) solved the radiation transfer in the vertical
direction, assuming that, the photons can escape only in the vertical
direction.  
For high accretion rate disk, however, the advected flow can trap
significant fraction of photons; the photons propagate not only in the
vertical direction but also in the radial direction. 
According to the preliminary results
 by Ohsuga, Mineshige, \& Watarai (2003), 
 the spectral hardening factor changes from 1.2 to about 2.3. 
This hardening factor depends on the disk model. 
Thus, the spectral hardening factor possibly exceeds 2.0. 
In recent study Shimura \& Manmoto (2002) also show that
 the general relativistic effect is important for solving the radiation
 transfer equations including the photon trapping. According to their
 results the hardening factor is more than $\kappa > 100$ in extreme
 Kerr hole and high mass accretion rate case ($\dot{m}$=100).   
We have not yet reached the consensus what the most appropriate value
is. 
Therefore, we treated the hardening factor as a parameter.

Here we adopt relatively large value $\kappa=2.0$ because of high
$\dot M$ cases being considered here and set the color temperature
 to be $\Tin = \kappa T_{\rm in}^0$.  
The effects of inclination angle and Doppler beaming are
ignored in the present fitting.  

\begin{figure} [h]
%\figurenum{3}
\epsscale{0.7}
\plotone{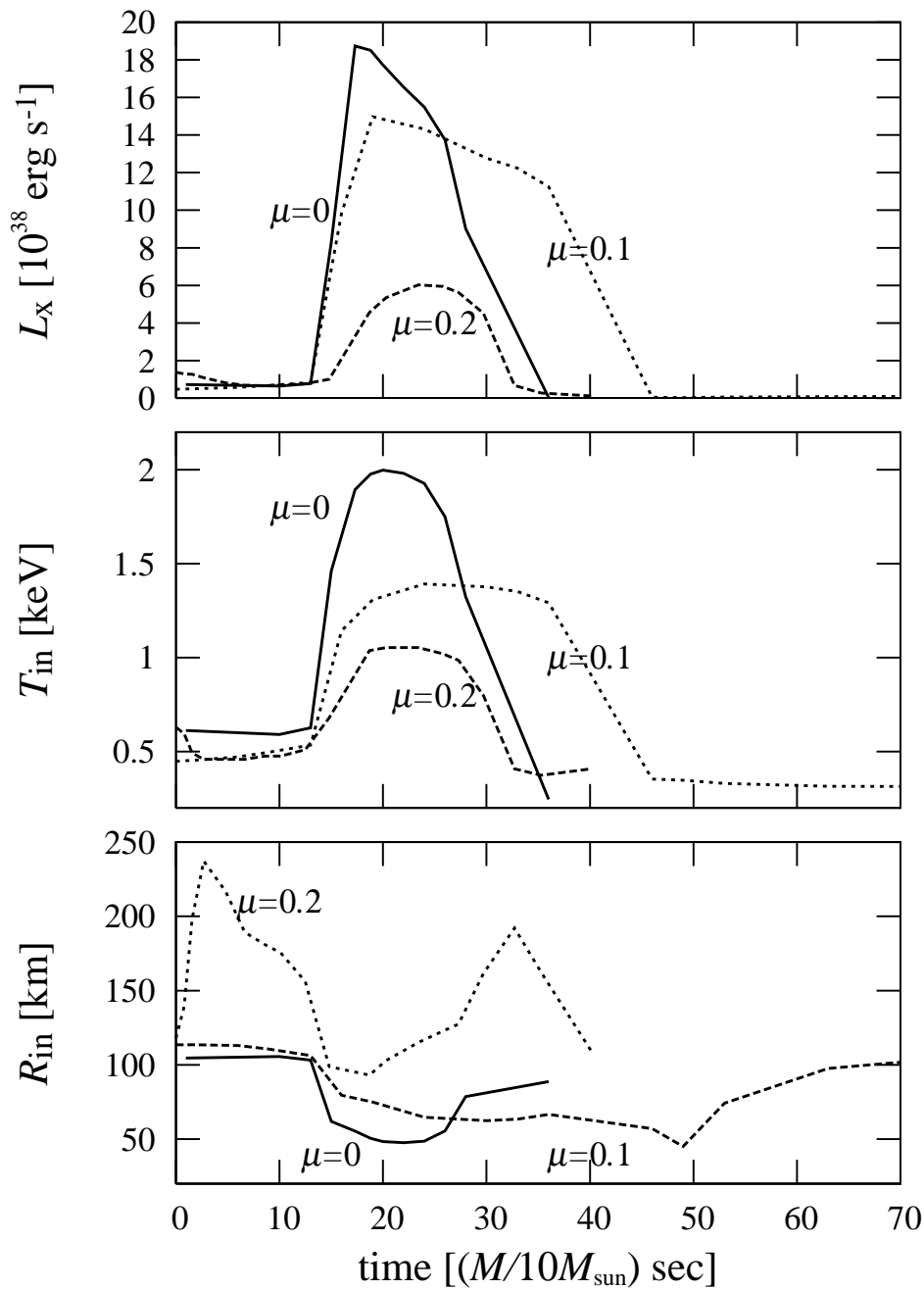}
%\plotone{ltinrin.ps}
\caption{Time variations of the X-ray luminosity, $L_{\rm X}$,
 the maximum temperature, $\Tin$,
 and the inner disk radius, $\Rin$, respectively. 
The horizontal axis is the elapsed time in sec. 
The solid, dashed, and dotted curves represent the cases
 with $\mu=0$, $\mu=0.1$, and $\mu=0.2$, respectively. 
 All the spectra are fitted in the energy range of 2.0-20 keV(for
burst phase) and 2.0-10 keV (for decay phase).
%0.2-8.0 keV (0.2-5.0 keV). 
}
\label{fig:fit}
\end{figure}

Figure \ref{fig:fit} displays the time variations of the X-ray
luminosity, maximum temperature ($\Tin$), 
and the inner edge radius of the disk ($\Rin$),
from the top to bottom, respectively (we also fitted the initial burst
for $\mu=0$ model). 

We calculated the X-ray luminosity by using $\Rin$ and $\Tin$, following
equation (4) in Makishima et al. (2000).    
In the rise phase 
the luminosity increases by about one order of magnitude, 
 $\Tin$ increases by a factor of $\sim$ 3, and
 $\Rin$ decreases by a factor of $\sim2$. 
At the end of the decay phase (corresponding to the time sequence 4)
we see a transient rise in $\Rin$. 
This is because in this phase
the temperature profile is rather flat up to $\sim 50 \rg$
so that the fitting is affected by the emission
 from this extended regions.
However, the transient rise may depend on
 the energy range used for fitting. 

%Finally,
%figure \ref{fig:fit2} summarizes the time variations of $\Tin$, $\Rin$, and
%$L_{\rm x}$ during the course of one burst cycle. 
%The maximum temperature increases and the inner disk radius decreases
%during the rise phase, while the opposite is true in the decay phase.
%{\bf We also calculate the case of inner boundary $\rin=2.5 \rg$, and
%fit the spectrum with MCD model. 
%The time variation of derived $\Rin$ parameters are small. }

We note that to see the significant variation of $\Rin$,
we should set the inner boundary of the disk as small as possible. 
Interestingly, however, evolutionary paths are different 
for the rise and decay phases in these diagrams.
We will discuss about the comparison between our model and the observed
objects in the next section.

\section{Discussion}

\subsection{Comparison with the observation of Microquasar GRS 1915+105}
We have calculated the time evolution of a near-critical accretion disk
undergoing the relaxation oscillations between the low mass accretion
rate branches and high mass accretion rate branches. 
We, especially, focus our study on the time changes of the apparent
inner-edge radius, $\Rin$, and the maximum temperature, $\Tin$, finding
peculiar behavior during the course of one burst cycle. 
In this subsection we discuss to what extent our model calculation can
explain the X-ray observations of GRS 1915+105.

\begin{figure} [h]
%\figurenum{5}
\epsscale{0.6}
\plotone{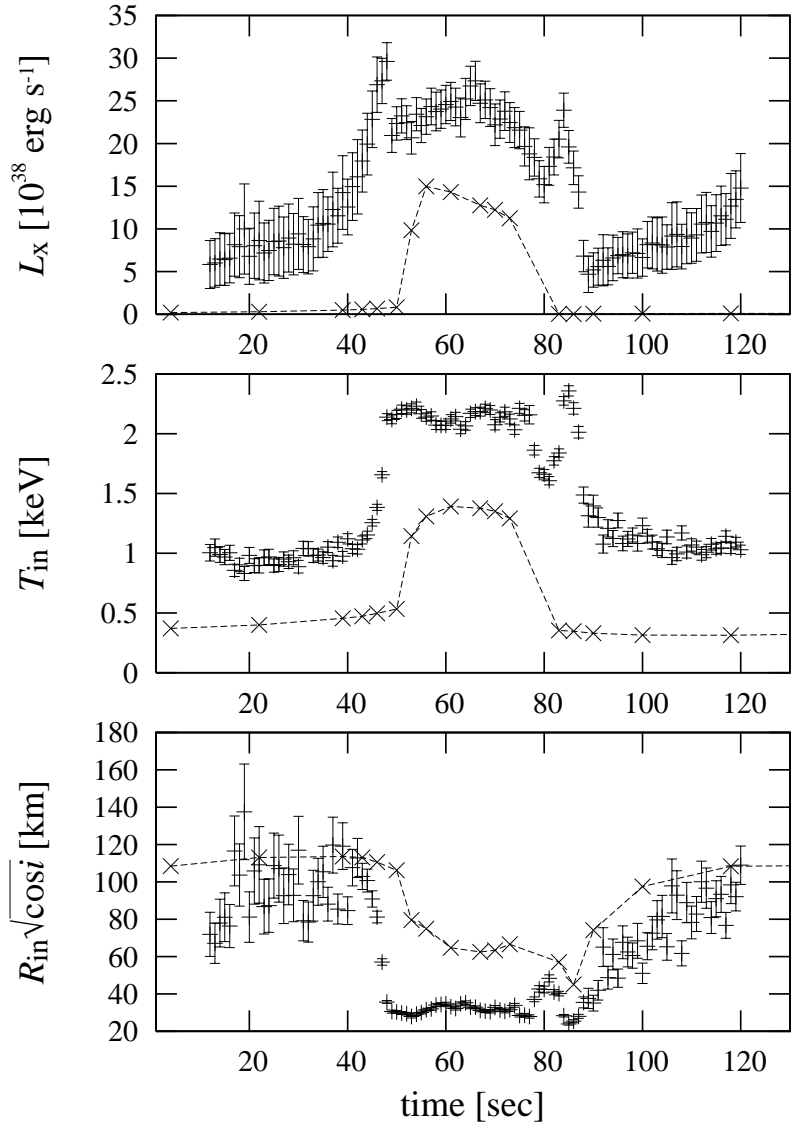}
%\plotone{yamaoka1.ps}
%{rg-2.2.dat/yamaoka2.ps}
\caption{The same as figure \ref{fig:fit} but from the observations 
of GRS 1915+105 (courtesy of K. Yamaoka).
 Cross plus dashed lines represent the derived fitting parameters on
 $\alpha=0.1, \mu=0.1$ model. 
 We assume the spectral hardening factor $\kappa=1.7$. 
 The inclination angle $i$ and the distance $D$ are 
 assumed to be $i=70$ and $D=12.5$ kpc (Mirabel \& Rodriguez 1994).  }
\label{fig:yamaoka1}
\end{figure}

Figure \ref{fig:yamaoka1} shows the results of the fitting to the
observational data of GRS 1915+105 made by Yamaoka et al. (2001) 
based on the MCD model. 
It is clearly seen in this figure
that the temperature increases and the inner radius 
decreases in the burst phase (see also Taam, Chen, \& Swank 1997).  
This is what we obtained in the calculations with $\mu = 0.1$,
thus we can conclude that qualitative features can be explained
by our model.

However, the precise fitting is not easy.
According to model calculations, the burst properties
can vary by changing the black-hole mass, $\alpha$ parameter,
or viscosity prescriptions.
Although the observed time changes of $\Tin$ and $\Rin$
agree well when $\mu = 0.1$,
the burst amplitude is too large to reconcile with the observation.
Model with $\mu = 0.2$, in contrast, can reproduce the observed amplitudes
of the luminosity variations during the burst, but the shape of the
light curve does not match the observation in the sense that
the burst rise is slower than that of the observation.
The observed time changes in $T_{\rm in}$ and $R_{\rm in}$ 
are not reproduced, neither.
That is, we cannot perfectly fit the observations by
simply changing the $\mu$ value.
Likewise, changing $\alpha$ values does not alter the results qualitatively,
but just results in changing the timescales. 
If we take smaller $\alpha$, say $\alpha=0.01$,
we will obtain longer burst duration, about 19 s, since 
the burst duration time is $\propto \alpha^{-0.64}$ (HMK91).

We should also remark that the absolute value of temperature is not
well reproduced, since $\Tin \sim 1 $keV in the observations,
whereas $\Tin \sim 0.6$ keV in our model, even if we adopt
a large spectral hardening factor of $\kappa \sim 2.0$.
 To fit the observations during the low-luminosity state,
 we need to require that
 there should be additional mechanisms to further
 harden the spectral so as to achieve $\kappa \sim 3$.
%This means, we need additional mechanism that causes a further
%spectral hardening.  
As mentioned Introduction, this issue
is common to other high-luminosity sources, such as ULXs.

There also exist other uncertainties arising due to
jet or wind, corona, Compton scattering effects, and so on.
Also, the determination of the inclination angle is very important.
Recently some authors argue that dissipation of energy
in disk corona or in jets can affect the disk evolution
%contribution of corona dissipation and of jet carrying 
%a luminosity-dependent fraction of energy 
(Nayakshin et al. 2000; Janiuk, Czerny, \& Siemignowska 2000). 
They propose the phenomenological model, in which they assume
the total gas energy in both disk and jet components to be conserved
and seek for the functional form,
 which matches the observed burst duration and luminosity. 
It is then found that that the fraction of the jet energy 
has to increase during the burst phase to match the observation.
However, there is no strong  physical basis on this suggestion.

If some dissipation processes like corona,
 jet or wind work around the peak,
 the amplitude of the burst get smaller, which might help.
In other words, we need to somehow modify the S-shaped thermal
equilibrium curves by changing $\alpha$ as a function of temperature,
by including dissipation processes in jets or corona at the peak, 
or something else.   
It might be interesting to note that the disk-instability model
for dwarf-nova outbursts has a similar problem; the constant $\alpha$
model cannot produce clear-cut outburst-quiescence light curves
but instead only exhibits small-amplitude fluctuations 
(see, e.g., figure 1 of Mineshige \& Osaki 1985).
To fit the observations, therefore,
one needs to increase $\alpha$ values in the hot state,
compared with that in the cool state, by a factor of $\sim 3$.
%Hence, to examine the jet or corona effect, 
%we have to treat them more carefully. 

\subsection{Is there a Kerr black hole?}

In the high mass-accretion-rate flow, the disk inner edge can be smaller
than $3 \rg$ even for a non-rotating black-hole (Watarai et al. 2000).
On the other hand, it is well known that
the case of a Kerr black-hole can also explain
the high disk temperature and small inner-disk radius, as are observed,
because of smaller radius of the marginally stable orbit.  
If the black hole is an extreme Kerr black hole, especially,
the disk inner edge should be as small as $0.5 \rg$.

Recently, mass estimation was made for the black hole of the microquasar
GRS 1915+105, which is $14\pm 4 M_{\odot}$ (Greiner et al. 2001).
 The corresponding Schwarzschild radius is 42$\pm 12$ km. 
If we take the face value derived by Yamaoka et al. (2001), the inner
radius of the disk should be $\Rin \ge \Rin \sqrt{{\rm cos} i}\sim 30$ km
during the burst, while $\Rin \sim 21$ km for the extreme Kerr hole. 
This result seems to rule out the extreme Kerr hole
interpretation at least for GRS 1915+105,
however, we should keep in mind that
we are only able to detect photons practically from the region outside
$\sim 2 \rg$ ($\rm \sim$ 82 km) owing to the significant photon
redshifts.  
It thus follows that it is not easy to distinguish 
whether there exists an extreme Kerr black hole or not.

At present, therefore, both possibilities (the Kerr hole 
and the super-critical accretion) are viable, but
we can at least rule out the extreme Kerr hole case, since
then $R_{\rm in}$ should stay constant at a small value,
not depending on the luminosity of the disk, contrary to the observations.

\subsection{Uniqueness of GRS 1915+105}

The greatest mystery involved with GRS 1915+105 resides in
why only this source exhibits various types of activities, including
quasi-periodic light oscillations.
High mass accretion rates may not be the only reason,
since other luminous sources, like GS 2023+338,
 whose luminosities seem to reach the Eddington luminosity
 (Tanaka 1989) did not record similar behavior.
Further, Kubota et al. (2001) analyzed the X-ray spectrum
 in some Galactic black hole candidates,
 revealing that XTE J1550--564,
 one of the most well-observed microquasars with RXTE,
 does not behave like GRS 1915+105 during the outburst,
 although the spectrum evolution is somewhat similar to
 that of GRS 1915+105;
 i.e.,  $\Tin$ increases and $\Rin$ decreases,
 as the luminosity increases in their burst phase.
They also indicate that the spectrum of XTE J1550--564 can be best
fit if we add the anomalous (Comptonized) component in the decay phase,
whereas no such component is needed in the rise phase.
This means, the disk structure differs in the rise and decay phases.
What discriminates these objects and other canonical black hole
X-ray transients?  
Do other bright ultra-luminous X-ray sources show bursting properties
similar to GRS 1915+105 but not observed yet?
These still remain unsolved problems.

It is interesting to note in this respect that
for relatively large viscous parameter, $\alpha \gtrsim 0.3$, 
a transition from the optically thick standard-disk branch 
to the optically thin ADAF occurs (Takeuchi \& Mineshige 1998).  
The transition timescale is less than a few sec 
for a stellar mass black-hole ($M_{\rm BH}=10M_{\odot}$). 
It might be that the solution is not unique and some hidden parameter
triggers the different behavior of near-critical accretion flow.

%Recently, the spectral transitions of GRS1915+105 from low/hard state to
%high soft state were observed by the $India X-ray Astronomy Experiment
%(IXAE)$ and RXTE (Rao et al. 2000; Naik et al. 2002). 
%This issue also our future work.

\section{Conclusions}

We calculated the evolution of the inner part of the high-luminosity
 accretion disk, carefully considering the transonic nature of the flow
in the innermost region.
We obtain the following results:

\noindent{1.} 
 Our calculations confirm the previous results
 that the state transitions between the standard-disk
 and slim-disk branches occur during the burst phase. 
Around the burst peak,
 the inner edge of the disk lies at $< 3\rg$
 and there are significant photons coming from
 the region inside $3 \rg$. 
%Moreover, the temperature profile of the disk shows the radial
% dependence, $T_{\rm eff} \propto r^{-1/2}$ during the burst. 
%These results are consistent with the analytical relation
% found for the steady model by Watarai \& Fukue (1999). 

\noindent{2.}
To inspect the X-ray observational features in GRS1915+105, 
we performed spectral fitting to the calculated spectra with
 the fitting parameters of $\Tin$ and $\Rin$. 
As to model with $\mu = 0.1$
 we find an abrupt transient increase in $\Tin$ 
and a temporary decrease in $\Rin$ during an outburst, 
both of which were actually observed in microquasar GRS 1915+105.
However, the amplitudes of light variations and
the absolute value of $\Tin$ are too
large to reconcile with the observational data.  
When $\mu=0.2$, in contrast,
the burst amplitude agrees with the observed one, but
the behavior of $\Tin$ and $\Rin$ differ from the observed one.
%the burst rise takes longer time ($\sim$ 7--10 s) than in 
%GRS 1915+105 ($\sim$ a few s).  
It is difficult to match the observation by simply varying $\alpha$ or $\mu$.
It is also important to note that the fitting results are sensitive to
the photon energy range used for the fitting.

\noindent{3.}
% The calculated burst duration does not perfectly reproduce the
% observation, but, if we take smaller $\alpha$, say $\alpha=0.01$,
% we will obtain longer burst duration, about 19 s, since 
%the burst duration time is $\propto \alpha^{-0.64}$ (HMK91).
To summarize,
although the precise fitting is left as future work, we can safely
 conclude that our ``limit-cycle oscillation'' model can explain the
basic observational features of GRS 1915+105.
It is then required that the spectral hardening factor
at high luminosities close to the Eddington luminosity
should be about 3 instead of less than 2 as is usually assumed.
This means, we need additional mechanism that causes a further
spectral hardening.

\acknowledgments

The authors thank the Yukawa Institute for Theoretical Physics at Kyoto
University. Discussions during the YITP workshop YITP-W-01-17 on ``Black
Holes, Gravitational Lens, and Gamma-Ray Bursts'' were useful to
complete this work. 
Especially, we would like to thank R. Narayan, C. Done, K. Yamaoka and
T. Kawaguchi for useful comments and discussions. 
Numerical computation in this work was carried out at the Yukawa
Institute Computer Facility. 
This work was supported in part by the Grants-in Aid of the
Ministry of Education, Science, Sports, and Culture of Japan
(14001680, KW).

%\appendix
%\section{Appendicial material}

\clearpage
\end{document}